\documentclass[twocolumn,pre]{revtex4}
\usepackage{graphicx}
\usepackage{times} 
\usepackage{palatino} 
\usepackage{newcent} 
\usepackage{bookman}

\newcommand{\bi}{\begin{itemize}}
\newcommand{\ei}{\end{itemize}}
\newcommand{\be}{\begin{equation}}
\newcommand{\ee}{\end{equation}}
\newcommand{\bea}{\begin{eqnarray}}
\newcommand{\eea}{\end{eqnarray}}
\newcommand{\ba}{\begin{array}}
\newcommand{\ea}{\end{array}}

\begin{document}
\title{Damage spreading in small world Ising models}
\author{Pontus Svenson$^a$\footnote{Current address: Dept of Information Fusion, Swedish Defense Research Agency, SE 172 90 Stockholm, Sweden} 
and Desmond A. Johnston$^b$}
\affiliation{
$^a$ tfkps@fy.chalmers.se \\
Institute for Theoretical Physics \\
Chalmers University of Technology and G\"oteborg University \\
SE-412 96, Gothenburg, Sweden \\
\\
$^b$
des@ma.hw.ac.uk \\
Department of Mathematics\\
Heriot-Watt University\\
Riccarton\\
Edinburgh, EH14 4AS, Scotland
}
\date{\today}

\begin{abstract}
We study damage-spreading in the ferromagnetic Ising model
on small world networks using Monte Carlo simulation with
Glauber dynamics. The damage spreading temperature $T_d$
is determined as a function of rewiring probability $p$
for small world networks obtained by rewiring the 
$2D$ square and $3D$ cubic lattices. We find that the 
damage for different values of $p$
collapse onto master curves when plotted against
a rescaled temperature and that the distance
between $T_d$ and the critical temperature $T_c$ increases
with $p$. We argue that when using the Ising model to study
social systems, it is necessary to place the spins on a
small world network rather than on a regular lattice.
\\\\
PACS numbers: 
75.10.Hk (Classical spin models),
75.40.Mg  (Numerical simulation studies),
75.10.Nr (Spin-glass and other random models)
\end{abstract}

\maketitle

\section{Introduction}

The Ising model is one of the most important models of
statistical mechanics. It and its generalisations
have been used to model a variety of 
natural phenomena, ranging from biology to
computer science and social science
(e.g., ~\cite{anderson,li:lectures,dickman,galam}).
For instance, many social systems can be modelled by letting 
spin up/down denote different opinions or preferences. In such models,
a ferromagnetic interaction is interpreted as two people who prefer to
agree, while an antiferromagnetic interactions means that they want to
disagree. A magnetic field adds a bias than can be interpreted as
``prejudices''  or ``stubbornness'', while the randomness induced by
a finite temperature can be seen as a ``free will''.

Damage spreading is a tool for studying the influence of perturbations
on the equilibrium state of a system. It has been used to
determine some properties of the energy landscape for
disordered spin systems~\cite{heeremaritort}, and also has great uses
for playing ``what if''-type scenarios in models of
complex systems. For a voter model, for instance,
damage spreading studies how much influence a (small) set of voters
can have over the final outcome of the election.
Damage spreading was first used by Kauffman~\cite{kauffman}
as a tool for studying biologically motivated dynamical systems,
but has since found widespread use also in physics (e.g.,~\cite{stanley87}).

Damage spreading works by duplicating an equilibrium spin
configuration of a system and changing a fraction $d_0$ of the
spins. Both systems are then subjected to the same thermal
noise and the distance between them is calculated.
In Monte Carlo simulations, both systems are simulated simultaneously:
the same spin is selected for spin-flip in both systems, and the
same random number (``thermal noise'') 
is used to determine whether an energy-raising
flip should be performed.

After equilibrating both systems, 
the Hamming distance (the number of different spins)
between the spin configurations $S^{\alpha}$
and $S^{\beta}$
\be
h(S^{\alpha}, S^{\beta}) = 
\frac{1}{N} \sum_i (1-\delta_{S^{\alpha}_i}^{S^{\beta}_i})
\ee
(where $\delta$ is the Kronecker delta function) is measured.
The Hamming
distance can also be expressed in terms of the Parisi overlap~\cite{parisi}
\be
q=\frac{1}{N}\sum_i S^{\alpha}_i S^{\beta}_i = 1 - 2 h  .
\ee

Most
of the work on both spin models and damage spreading
place the spins either
on a finite-dimensional lattice or on a random graph.
Here we instead use 
small world graphs~\cite{wattsstrogatz:nature,wattsbook}
to study the ferromagnetic
Ising model on graphs interpolating between 2 and 3-dimensional
simple cubic lattices and random graphs with the same connectivity.
The Hamiltonian of our model is
\be
H = - \sum_{i<j} J_{ij} S_i S_j
\ee
where $J_{ij}$ is 1 if and only if there is an edge between spins
$i$ and $j$ and 0 otherwise.

Small world graphs are intermediates between a regular lattice and 
a random graph; they have previously been
used to study, e.g., computation, diffusion, and spreading of diseases.
The original motivation for studying small worlds is that they
possess both small diameters (like a random graph~\cite{bollobas}) and
a high degree of clustering (like a regular lattice).
For examples of real world networks with small world
characteristics and reviews of previous work, see, e.g., 
\cite{wattsbook,newmanreview,albertreview,dorogovtsevreview}. 

The small world is constructed 
by considering in turn all the edges $(i,j)$ of a lattice and
with some probability $p$ replacing it with a random edge $(i,k)$.
The rewiring parameter
$p$ thus determines how many of the links are removed and can be used to 
interpolate between the regular lattice and a random graph.
Note that the small world for $p=1$ differs slightly from a 
random graph, since all nodes are guaranteed to have a local
connectivity of at least $z/2$ where $z$ is the connectivity of the
regular lattice. The distribution of connectivities is more broad
for the small world with $p=1$. We chose to use the small world model
where links are rewired and not the one where they are added because
we wanted to keep the average connectivity of the graphs the
same for all $p$.

The use of small world graphs to study physical
models has so far been limited. 
Barrat and Weigt~\cite{barratweigt} and Gitterman~\cite{gitterman}
have used them
to study the crossover from $1D$ to mean-field behaviour
for the
ferromagnetic Ising model, finding a disorder-order
transition at a finite temperature $T_c(p)$ 
for any $p>0$, provided that the system size is large enough.

%Small world graphs is one alternative to random graphs
%and regular lattices; there are also others. A recent 
%study~\cite{newmanstrogatzwatts2000}
%of a wide variety of 
%collaboration graphs in science and business as well as graphs
%of Internet 
%connections have shown
%that some of them can be described using random graphs but with a 
%different distribution
%of degrees of the nodes in the graph; these graphs may or may not
%also be clustered enough to be small world networks.

Most of the work on small world networks has started by
rewiring a one-dimensional
ring lattice, but here we instead use the $2D$
square and $3D$ simple cubic lattices. One reason
for doing this is that while the $1D$ Ising model is trivial
and disordered for all finite temperatures, the $2D$ and $3D$ versions 
are ordered below a critical temperature $T_c$. The $2D$ model
can be solved exactly, while for $4D$ and
higher-dimensions, mean field theory explains
the phase transition (see, e.g.,~\cite{binney}).
An important concept in the study of phase transitions and
critical phenomena is that of universality class. 
Models displaying the same behaviour close to $T_c$
are said to be in the same universality class, and it
turns out that there are many fewer universality classes
than models. 
Putting spin models on small world graphs provides an opportunity
to study the crossover from a finite-dimensional universality class
to mean field behaviour. Here we restrict ourselves to determining
$T_c$, but it would also be interesting to see how the critical
exponents change as $p$ is increased.

It should be noted that the small world networks used here differ from
those obtained by rewiring a ring lattice in one respect:
their clustering coefficient does not display the same
threshold behaviour as a function of $p$: it
starts at 0 for $p=0$ (since the regular lattices
used are bipartite) and then grows to the random graph value.
The graphs used here are however still clustered in the sense that
if $j$ and $k$ are neighbours of $i$, then there is a short path
between them that does not pass through $i$.

While the emphasis in the present work is on the damage
spreading behaviour of the model, we also determined the
critical temperature $T_c$ for the order-disorder transition.
This was done primarily in order to compare it with the
damage spreading temperature $T_d$; the numerical accuracy
of $T_c$ is smaller than that for $T_d$.

The Monte Carlo method used was the standard single spin-flip
Metropolis~\cite{metropolis}
algorithm. In each time-step, $N$ spin flips
are attempted. For each flip-attempt, a spin is randomly
selected and the energy-change $\Delta H$
if it is flipped is calculated.
If the change in energy is negative, the spin is always flipped,
 otherwise it is flipped with probability
$e^{-\Delta H/T}$ where $T$ is the temperature.
We also did some runs using different MC procedures 
(heat-bath algorithm, spin-exchange, using an ordered update
instead of a random). We found that using the heat-bath
algorithm caused the damage to heal
at temperatures close to and above $T_c$,
while for the spin-exchange
dynamics with the Metropolis algorithm the damage spreads
for all temperatures. Updating the spins in order instead of
randomly gives a smaller damage for all temperatures.
These results agree with the results of 
Vojta~\cite{vojta:exchangedamage,vojtaschreiber,vojta:1997}
for the standard Ising model.

In most of the simulations, we used the Mitchell-Moore
additive random number generator (see, e.g.,~\cite{knuth2}
for a description). We also did some runs with the standard
C library's {\tt drand48()} generator and found the same
behaviour. All simulations were averaged over $N_l$
different rewiring procedures, and for each small world graph
an average over $N_r$ independent Monte Carlo runs was performed.
Typical values were $N_l = N_r = 10$, but this was varied
for some runs in order to check self-averaging. No significant differences
in behaviour was found.

Our simulation procedure was simple. After equilibrating
the system (using simulated annealing),
a copy is made and $d_0 N$ spins in it are flipped. Both
systems are then simulated using the same random numbers to
determine which spin to select and whether or not to flip it. After
equilibrium has been reached again, we start measuring the damage as
well as other quantities such as the magnetisation and energy and
their standard deviations. We used $d_0 = 0.01$ in all of the
simulations presented here; none of the results
presented are sensitive to the exact value of $d_0$.
In order to check the dependence on initial conditions,
we also performed some runs damaging a non-equilibrated system;
these gave the same results.

\begin{figure}
\centering
\leavevmode
\includegraphics[width=.75 \columnwidth]{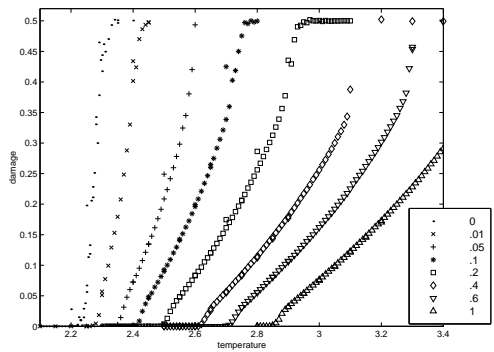}
\vspace*{2mm}
\caption{This figure shows the damage as a function of temperature
for small world graphs obtained by rewiring
a $100\times 100$ $2D$ square lattice with (from left to right) 
$p=0$, 0.01, 0.05, 0.1, 0.2, 0.4, 0.6, and 1.
For each $p$, an average over 10 graphs and 10 restarts per graph
was performed. The location of $T_d$ shifts to higher temperatures as $p$
is increased, and the slope of $d(T)$ decreases.}
\label{2ddamtfig}
\end{figure}

Figure~\ref{2ddamtfig} shows the end-damage as a function of
temperature for $p$ ranging from 0 to 1. The rewired lattice in
this figure is the $2D$ square with $N=10^4$ spins. We tested
some different system sizes and found that this seems to be a
large enough number of spins that finite-size effects are minimised.
The data was averaged
over $N_l=10$ graphs and for each graph the Monte Carlo simulation was
restarted $N_r=10$ times in order to improve numerical accuracy. 
Error bars for the damage in this and the following
figures were determined to be at most on the order of 0.01 and
in almost all cases considerably smaller. Note though that the errors
increase with $p$, as should be expected since the averaging
becomes more important for large $p$. 

\begin{figure}
\centering
\leavevmode
\includegraphics[width=.75 \columnwidth]{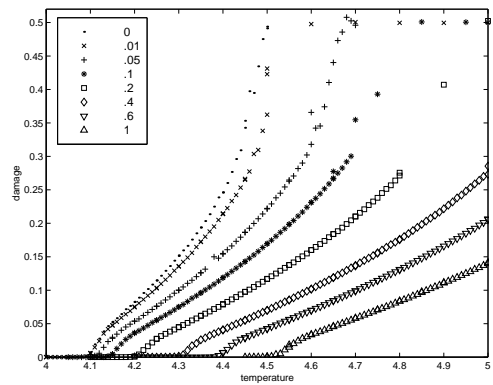}
\vspace*{2mm}
\caption{Here we show the damage as a function of temperature
for small world graphs obtained by rewiring
a $20\times 20\times 20$ $3D$ cubic lattice with (from left to right) 
$p=0$, 0.01, 0.05, 0.1, 0.2, 0.4, 0.6, and 1.
For each $p$, an average over 10 graphs and 10 restarts per graph
was performed. As in the $2d$ case, the location of 
$T_d$ shifts to higher temperatures as $p$
is increased, and the slope of $d(T)$ decreases.}
\label{3damage}
\end{figure}

Figure~\ref{3damage} shows the corresponding data for the $3D$ lattice. The
system size here is $N=8000$ and $N_l=N_r=10$ as for the $2D$ data.

\begin{figure}
\centering
\leavevmode
\includegraphics[width=.75 \columnwidth]{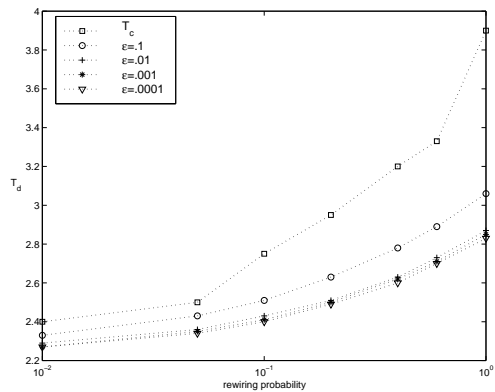}
\vspace*{2mm}
\caption{For the same data as in figure~\ref{2ddamtfig}, this figure
shows the $p$-dependence of $T_c$ (squares) and $T_d$ for some different
$\epsilon$. Note the logarithmic scale of the $p$-axis in this plot.
It is clear that $T_d$ is independent of $\epsilon$
for small enough $\epsilon$'s.}
\label{2ddifferenteps}
\end{figure}

We can define a damage spreading temperature $T_d(\epsilon)$ as the lowest
temperature for which the damage
$d$ is larger than some (small) $\epsilon$,
\be
T_d(\epsilon) = \min \{ T : d(T) > \epsilon \} .
\label{tdeqn}
\ee
In the limit as $\epsilon \to 0$, our $T_d(\epsilon)$ converges
to the standard $T_d$ which is defined as the lowest
temperature for which the damage is non-zero. We use
a non-zero $\epsilon$ in equation~\ref{tdeqn} when determining
$T_d$ from our data becuase
using a $\epsilon$ smaller than the error-bar for the damage
would lead to noise in $T_d$.
Figures~\ref{2ddamtfig} and~\ref{3damage}
show clearly that $T_d$ increases with $p$,
as is to be expected. In order to quantify this, figure~\ref{2ddifferenteps}
compares $T_d$ to the order-disorder transition temperature $T_c$
for the $2D$ data. 
The figure shows $T_d$ for $\epsilon=10^{-4}$, $10^{-3}$, $10^{-2}$,
and $10^{-1}$; it is clear that the definition of $T_d$ is independent
of $\epsilon$ for small enough $\epsilon$'s. The temperature
where the damage attained its maximum value of $0.5$ seems
to approach $T_c$; this is in agreement with previous
work~\cite{hinrichsen}.
The critical temperature  $T_c$ 
was determined as the temperature at which the Binder's cumulant
\be
c = \frac{\langle m^4 \rangle}{\langle m^2 \rangle^2}
\ee
curves for large system sizes cross.
For the 2D lattice, the largest system simulated consisted of
$10^4$ spins, while in the 3D case shown in 
figure~\ref{3depscomparison} below, system sizes up to $21^3=9261$ were used
to determine $T_c$.
The error bars for $T_c$ are larger than for $T_d$; note that the
mean-field value for (regular)
random graphs with coordination number $z$ is $T_c=z$.
The value of
$T_d$ obtained for $p=1$ here is in reasonable agreement to the one for
normal random graphs.

%for random graphs, $t_d=0.8 \gamma$ extremely good fit for $\gamma \geq 3$.

\begin{figure}
\centering
\leavevmode
\includegraphics[width=.75 \columnwidth]{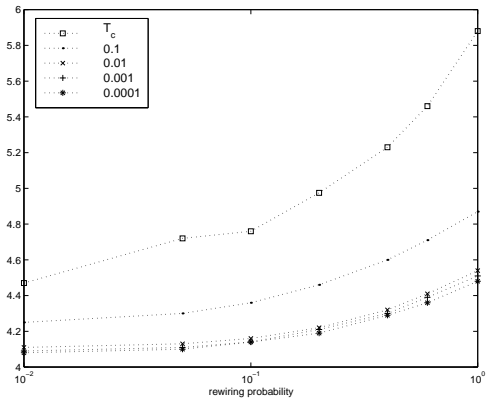}
\vspace*{2mm}
\caption{Here we show $T_c$ (squares) and 
$T_d$ as a function of $p$ for some different
$\epsilon$ for the $3d$ case.
Here, too, the values for $T_d$ are independent
of the exact value of $\epsilon$, provided that it is small enough.}
\label{3depscomparison}
\end{figure}

\begin{table}[h]
\centering
\begin{tabular}{|c|c|c|c|c|c|c|c|c|c|}
\hline
$2D$ & p     & 0 & 0.01 & 0.05 & 0.1 & 0.2 & 0.4 & 0.6 & 1 \\
\hline
 & $T_d$ & 2.24 &2.28 &2.34 & 2.40& 2.49& 2.60 &   2.70 &2.83 \\
\hline
$3D$ & p     & 0 & 0.01 & 0.05 & 0.1 & 0.2 & 0.4 & 0.6 & 1 \\
\hline
 & $T_d$ &  4.08 &4.08 &4.10 &4.14 &4.19 &4.29 &4.36& 4.48 \\
\hline
\end{tabular}
\vspace*{5mm}
\caption{$T_d$ for the small world starting from a $2D$ and $3D$ lattices.}
\label{tdtab}
\end{table}

Table~\ref{tdtab} shows the values for $T_d$
for different $p$
for small worlds obtained by rewiring the $2D$ square and $3D$
cubic lattices.
For $p=0$, we get values in agreement with those reported in the 
literature\cite{grassberger95a,grassberger95b,vojtaschreiber}.
%3d: $T_s=4.162$ for 309x309x310 sites, helical boundary conditions
%and checkerboard update(grassberger\cite{grassberger95a,grassberger95b},
%as cited in~\cite{vojtaschreiber}. 
%2: somebody got 2.24??? grassberger\cite{grassberger95a} 
%gets $0.992 T_c = 2.25$???
%Interestingly, by using $\epsilon=0.4$ it is possible to to make the
%curves for $T_c$ and $T_d(\epsilon)$ almost overlap.

\begin{figure}
\centering
\leavevmode
\includegraphics[width=.75 \columnwidth]{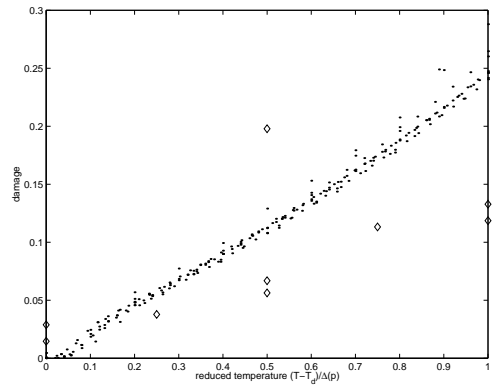}
\vspace*{2mm}
\caption{Same data as in figure~\ref{2ddamtfig}, plotted as a function 
of a  rescaled temperature.By plotting the damage as a function of 
a reduced temperature $\tilde{T}=(T-T_d)/\Delta(p)$, it is
possible to get collapse for all $p$ except $p=0$ (shown as small
squares in the figure),
which does not follow the same functional form as
the other curves.}
\label{2dcollapse}
\end{figure}

Scaling plots are used to combine data from runs with
different values of some parameter into one curve. 
In our case, we can make
the data for different $p$ fall onto the same curve by
plotting the damage as a function of a rescaled temperature
\be
\tilde{T} = \frac{T-T_d}{\Delta(p)} .
\label{tildeeq}
\ee
Our scaling ansatz is that the damage can be written as
\be
D(T,p) = f(\frac{T-T_d(p)}{\Delta(p)})   ,
\label{scalingeq}
\ee
for some $f$ which is independent of $p$.
In equation~\ref{scalingeq}, $\Delta(p)$ is determined by the
inverse of the rate
at which the damage develops for different $p$
\be
\frac{d D}{d T}(T=T_d) = 
\frac{1}{\Delta(p)} \frac{d f}{d \tilde{T}}(\tilde{T}=0)   .
\ee
$\Delta$ is an
increasing function of $p$; physically it tells us how much
more we must increase the temperature in order to get the
same increase in damage for different $p$:
\be
\Delta T \propto \Delta(p) \Delta D  .
\ee
The values for $\Delta(p)$ determined from the data in
figures~\ref{2ddamtfig} and~\ref{3damage} are shown in table~\ref{deltatab}.
We found a reasonable scaling $\Delta(p) \sim p^{\alpha}$
with $\alpha \approx 0.35$ for the $2D$ data and $\alpha \approx 0.2$
for the $3D$ data. The function $f$ turns out to be linear.

\begin{table}[h]
\centering
\begin{tabular}{|c|c|c|c|c|c|c|c|c|c|}
\hline
$2D$ & p     & 0 & 0.01 & 0.05 & 0.1 & 0.2 & 0.4 & 0.6 & 1 \\
\hline
 & $\Delta(p)$& 0.07& 0.1 & 0.18  & 0.25 & 0.3 & 0.38 & 0.43 & 0.5 \\
\hline
$3D$ & p     & 0 & 0.01 & 0.05 & 0.1 & 0.2 & 0.4 & 0.6 & 1 \\
\hline
 & $\Delta(p)$& 0.28& 0.3 & 0.36& 0.41& 0.5& 0.55& 0.61 & 0.7 \\
\hline
\end{tabular}
\vspace*{5mm}
\caption{$\Delta(p)$ for the $2D$ and $3D$ rewired lattices.}
\label{deltatab}
\end{table}

Figure~\ref{2dcollapse} plots the damage as a function
of $\tilde{T}$ for the $2D$ case. A very good
collapse is obtained for all $p>0$. 
%sometimes, it is possible to obtain data collapse by plotting curves
%representing different parameters. as can be seen in figure~\ref{2dcollapse},
%this applies here too for $p>0$.. values for $\Delta$:
%.07 .1 .18  .25  .3 .38  .43 .5 (2d values)
%values for $\Delta(p)$: .28 .3 .36 .41  .5 .55 .61  .7 (3d)
The data for $p=0$ can not be made to fall onto the same curve.
Note that
the distance between the master
curve and the $p=0$ data is larger than the estimated error bars.

\begin{figure}
\centering
\leavevmode
\includegraphics[width=.75 \columnwidth]{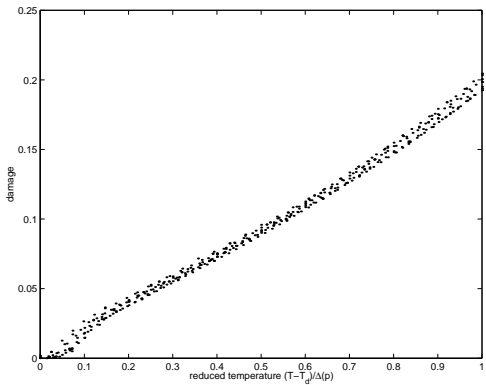}
\vspace*{2mm}
\caption{In contrast to the $2D$ case,
by plotting the damage as a function of 
a reduced temperature $(T-T_d)/\Delta(p)$, it is
possible to get collapse for all $p$ for the $3D$ data.}
\label{3dcollapse}
\end{figure}

Figure~\ref{3dcollapse} shows that, in contrast to the 2D case,
the $3D$ data do collapse onto one curve for all $p$, including
the $p=0$ (i.e., simple cubic lattice) case.

This shows some qualitative differences between the 2 and 
3-dimensional lattices. The way that damage spreads in the model can be
seen as a form of generalised random walk; we speculate that the difference
between the $2D$ $p=0$ data and the other data might be related to 
the differences (in, e.g., return time)
between random walks on $2D$ and $3D$/random
lattices~\cite{grimmettstirzaker}.

We also studied the approach to equilibrium of the damaged system.
Figure~\ref{2drelax} below shows the relaxation of the damage
as a function of the number of complete Monte Carlo sweeps
after the damage is introduced. The figure shows data for
$2D$ model with $p=0.4$; the relaxation behaviour for other
values of $p$ as well as for the $3D$ case is similar.
It is clearly seen that there
is a power-law for a short interval above $T_d$. 

\begin{figure}
\centering
\leavevmode
\centerline{\includegraphics[width=.8 \columnwidth]{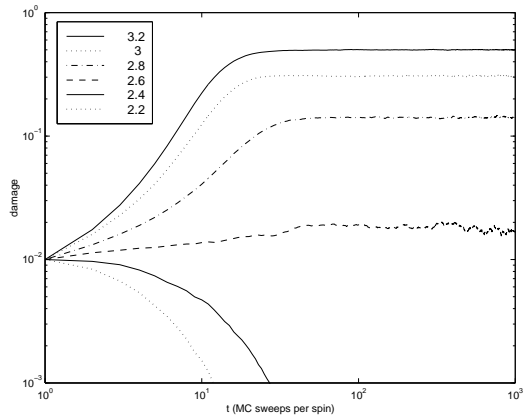}}
\vspace{1cm}
\caption{This figure shows the time-dependence of the damage for 
the $2D$ model with $p=0.4$ and $T=2.2$, 2.4, 2.6, 2.8, and 3.0.
The relaxation is exponential below $T_d$, and displays a power-law
for a short interval for $T>T_d$. The damage spreading transition 
takes place at $T_d\approx 2.6$.}
\label{2drelax}
\end{figure}

The data can be very approximately fitted to a form $d(t)\sim t^{a}$
with $a\approx 1.5 \pm 0.1$ for $T$ considerably larger than $T_d$
and for all $p>0$. The exponent for $p=0$ is significantly different,
$a\approx1.1$.

In conclusion, we found that the damage for different small worlds
fall onto a universal curve when plotted as a function of a
rescaled temperature.
The distance between $T_d$ and $T_c$ increases
as a function of rewiring probability
$p$, i.e., the range in temperature where the
model is ordered but small perturbations are important increases. 
This is important for models of social systems, where we can interpret
the temperature as a form of (random) ``free will''.

%for 3d ising model, $T_c = 4.5115$ and $T_d=4.162$.
%for 3d $\pm J$ model, $T_g=1.1$.

We believe that putting spin models on small world graphs provides
an ideal method not only of studying social models more realistically
but also of testing hypotheses regarding spin models. For instance,
it is an interesting open question how to accurately describe
the ground state and low-lying excitations of the $3D$ $\pm J$
spin glass model. By putting this model on a small world graph
and studying the crossover to the $p=1$ mean-field behaviour, it might
be possible to learn more about this. 
 
{\bf Acknowledgements:}
We would like to thank an anonymous referee for their comments.
P.S. thanks the Maths Department at Heriot-Watt and the
Edinburgh Parallel Computer Centre for hospitality and
the European Commission (grant number HPRI-CT-1999-00026)
for financial support.

\end{document}